\documentclass[prd,onecolumn,nopacs,nofootinbib]{revtex4}

\usepackage{amsmath}
\usepackage{graphicx}
\usepackage{amsfonts}
\usepackage{latexsym}
\usepackage{bbold}
\usepackage{wasysym}
\usepackage{calligra}
\usepackage{float}
\usepackage{ulem}
\usepackage{inputenc}
\usepackage{xspace}
\usepackage{url}
\usepackage{epstopdf}
\usepackage{tikz}

\newcommand\Lie{\pounds}

\newtheorem{mydef}{Definition}

\newcommand{\be}{\begin{equation}}
\newcommand{\ee}{\end{equation}}
\newcommand{\beq} {\begin{equation}}
\newcommand{\eeq} {\end{equation}}
\newcommand{\ba}{\begin{eqnarray}}
\newcommand{\ea}{\end{eqnarray}}
\newcommand{\w}{\wedge}
\newcommand{\vtt}{\vartheta}
\newcommand{\rf}{\rfloor}

\begin{document}

	\title{The Perfect Hyperfluid of Metric-Affine Gravity: The Foundation}
	
\author{Damianos Iosifidis}
\affiliation{Institute of Theoretical Physics, Department of Physics
	Aristotle University of Thessaloniki, 54124 Thessaloniki, Greece}
\email{diosifid@auth.gr}
	
	\date{\today}
	\begin{abstract}
We set the foundation and formulate the Perfect (Ideal) Hyperfluid. The latter represents the natural generalization of the usual perfect fluid structure where now  the microscopic characteristics of matter (spin, shear, dilation) are also taken into account, sourcing a non-Riemannian arena (i.e spacetime torsion and non-metricity) for Metric-Affine Gravity. We derive the energy tensors of this Hyperfluid structure and subsequently present the conservation laws obeyed by them. Finally, we consider a Cosmological application of this Perfect Hyperfluid  and classify some possible forms of this fluid structure.
		
	\end{abstract}
	
	\maketitle
	
	\allowdisplaybreaks
	
	
	\tableofcontents
	
	\section{Introduction}
	
	The Perfect Fluid notion, that we are acquainted with from GR, (or even metric extensions of the latter) has very broad and important applications. However, this  collective fluid representation (the Classical Perfect Fluid) is applicable only to matter with no internal structure. At the same time, it is known that in order to probe the non-Riemannian characteristics (torsion and non-metricity) of the underlying spacetime manifold, matter with intrinsic hypermomentum has to be used \cite{puetzfeld2008probing}. It is therefore natural to ask what would be the generalization-extension of the classical Perfect Fluid notion of GR, to Metric-Affine Gravity\cite{hehl1995metric,iosifidis2019metric}? In this direction, many interesting models of fluids with intrinsic hypermomentum have been considered in the past almost $20$ years ago (see \cite{obukhov1993hyperfluid,obukhov1996model,babourova1998perfect}). However, a generic formulation of an isotropic hyperfluid which represents the direct generalization of the Perfect Fluid was missing so far. Such an attempt was considered recently in \cite{damianos2020cosmological}. However, that model  was tailored only for Cosmological settings  and was also laid on the assumption of equivalence between the canonical and the metrical energy momentum tensors of matter. It is the purpose of this note to further extend this consideration by dropping the latter postulate along with homogeneity assumption in order to formulate a generalized isotropic hyperfluid.
	
	The paper is organized as follows. Firstly, setup our conventions and briefly discuss the basic concepts of a non-Riemannian Geometry\cite{eisenhart2012non}. Then we touch upon the sources of MAG, being the canonical energy momentum, the metrical energy momentum and the hypermomentum tensor, and also present their associated conservation laws. Subsequently we formulate the concept of the Perfect (Ideal) Hyperfluid by first giving its physical definition and later using the appropriate mathematical formulation in order to extract its energy tensors by demanding spatial isotropy. Continuing, we apply the conservation laws of MAG to the energy tensors we derived establishing, therefore, a complete description of the novel fluid model we propose. Finally, we consider the Cosmological application of our Hyperfluid model and classify the possible Perfect Cosmological Hyperfluids one can have depending on the equations of state among the hyperfluid variables.

\section{The Setup}
	
	We shall now briefly discuss some basic aspects of non-Riemannian geometry that are essential for our analysis. We will use the conventions of \cite{damianos2020cosmological,iosifidis2019metric} so we will go through them rather quickly here. 
	
Let us first fix notation. We shall use  letters from the Latin alphabet $a,b,c,...$ to denote anholonomic indices and Greek $\mu,\nu, \rho,...$ for holonomic ones (coordinate) both ranging from $0,1,2,...n-1$ where $n$ is the spacetime dimension. We consider an $n-dim$ non-Riemannian manifold endowed with a metric along with a linear connection $(\mathcal{M},g,\nabla)$. As usual, at each point $p$ on the manifold we can define a set of local frames $\{e_{a}\}$ spanning the tangent vector space $\mathcal{T}_{p}\mathcal{M}$ of the manifold at that point. We also define the coframe one forms $\{\vtt^{a}\}$, living in the cotangent space $\mathcal{T}_{p}^{\star}\mathcal{M}$, through the duality relation
\beq
e_{a} \rf \vtt^{b}=\delta_{a}^{b}
\eeq
where $\rf$ denotes the interior product. In addition, we assume the existence of a  $GL(n,R)$-valued linear connection $1$-form $\Gamma^{a}_{\;\;b}$ allowing us to define the gauge exterior gauge covariant derivative on arbitrary tensor valued forms.  From the gauge potentials $(g_{ab}, \vtt^{a},\Gamma^{a}_{\;\; b})$ we construct their associated field strengths corresponding to non-metricity, torsion and curvature, according to \cite{hehl1995metric}
\beq
\mathcal{Q}_{ab}:=-Dg_{ab}
\eeq
\beq
\mathcal{T}^{a}:=D\vtt^{a}=d\vtt^{a}+\Gamma^{a}_{\;\; b}\w \vtt^{b}
\eeq
\beq
R^{a}_{\;\; b}:=d\Gamma^{a}_{\;\; b}+\Gamma^{a}_{\;\; c}\w \Gamma^{c}_{\;\; b}
\eeq
respectively. In the above $D$ represents the exterior gauge covariant derivative, and obviously the non-metricity is an $1$-form while torsion and curvature are both $2$-forms. Also, we define the invariant volume element as $\mu:=\frac{1}{n!}\epsilon_{a_{1}...a_{n}}\vtt^{a_{1}}\w...\w \vtt^{a_{n}}$ and the Hodge dual for an arbitrary $p$-form $\Psi$ as $\star \Psi :=\frac{1}{(n-p)!p!}g^{a_{1}c_{1}}...g^{a_{p}c_{p}}\epsilon_{a_{1}...a_{p}b_{1}...b_{n-p}} \Psi_{c_{1}...c_{p}}\vartheta^{b_{1}}\wedge ...\wedge \vartheta^{b_{n-p}}$. Now, considering an appropriate local gauge transformation we can set the gauge such that\footnote{See \cite{hehl1995metric} for more details. Note that the conventions we are using here have some slight variations from the ones found there. }
\beq
\partial_{\nu}e_{\mu}^{\;\; a}-\Gamma^{\rho}_{\;\;\mu\nu}e_{\rho}^{\;\; a}+\Gamma^{a}_{\;\; b \nu}e_{\mu}^{\;\; b}=0
\eeq
where the expansions $e_{a}=e^{\mu}_{\;\; a}{\partial}_{\mu}$, $\vtt^{a}=e_{\mu}^{\;\; a}dx^{\mu}$, $\Gamma^{a}_{\;\;b}=\Gamma^{a}_{\;\;b\mu}dx^{\mu}$ along with the identity $e_{\mu}^{\;\; a}e^{\mu}_{\;\; b}=\delta_{a}^{b}$ , have been used.
The latter relation enables one to switch over to a holonomic description based on a metric $g_{\mu\nu}$ and an independent  affine connection $\Gamma^{\lambda}_{\;\;\;\mu\nu}$. In this instance our definitions for non-metricity, torsion and curvature respectively read\footnote{In the transition from the anholonomic to the holonomic frame description we have denoted $\mathcal{T}^{c}_{\;\;\mu\nu}:=-2 S_{\mu\nu}^{\;\;\; c}$ and $\mathcal{Q}_{abc}=Q_{bca}$.  Note also that the form indices appear on the very right of the component expansion of each object, for instance $\mathcal{Q}_{ab}:= \mathcal{Q}_{abc}\vtt^{c}$ and $\mathcal{T}^{a}=\frac{1}{2}\mathcal{T}^{a}_{\;\; bc}\vtt^{a}\w \vtt^{c}$. }
\beq
Q_{\alpha\mu\nu}=-\nabla_{\alpha}g_{\mu\nu}
\eeq
\beq
S_{\mu\nu}^{\;\;\;\lambda}:=\Gamma^{\lambda}_{\;\;\;[\mu\nu]}
\eeq
\begin{equation}
R^{\mu}_{\;\;\;\nu\alpha\beta}:= 2\partial_{[\alpha}\Gamma^{\mu}_{\;\;\;|\nu|\beta]}+2\Gamma^{\mu}_{\;\;\;\rho[\alpha}\Gamma^{\rho}_{\;\;\;|\nu|\beta]}
\end{equation}
Our definitions for the associated torsion and non-metricity vectors are
	\beq
S_{\mu}:=S_{\mu\lambda}^{\;\;\;\;\lambda} \;\;, \;\;\;
\tilde{S}_{\mu}:=\epsilon_{\mu\alpha\beta\gamma}S^{\alpha\beta\gamma} 
\eeq
	\beq
Q_{\alpha}:=Q_{\alpha\mu\nu}g^{\mu\nu}\;,\;\; \tilde{Q}_{\nu}=Q_{\alpha\mu\nu}g^{\alpha\mu}
\eeq
respectively. In addition, without the use of any metric, from the Riemann tensor we can derive the two contractions
\beq
R_{\nu\beta}:=R^{\mu}_{\;\;\;\nu\mu\beta}\;,\; \widehat{R}_{\alpha\beta}:=R^{\mu}_{\;\;\;\mu\alpha\beta}
\eeq
the first one defines, as usual, the Ricci tensor (which is not symmetric now) and the second one is the homothetic curvature. If a metric is given another contraction can be formed which reads
\beq
\breve{R}^{\lambda}_{\;\;\kappa}:=R^{\lambda}_{\;\;\mu\nu\kappa}g^{\mu\nu}
\eeq
 Note however that the Ricci scalar is always uniquely defined since
 \beq
R:=R_{\mu\nu}g^{\mu\nu}=- \breve{R}_{\mu\nu}g^{\mu\nu} \;\;, \;\;\; \widehat{R}_{\alpha\beta}g^{\mu\nu}=0
 \eeq
Finally, the affine connection can always be decomposed into a Riemannian piece (Levi-Civita connection) plus post Riemannian contributions, according to \cite{schouten1954ricci,iosifidis2019metric} 	
	\beq
	\Gamma^{\lambda}_{\;\;\;\mu\nu}=\widetilde{\Gamma}^{\lambda}_{\;\;\;\mu\nu}+N^{\lambda}_{\;\;\;\mu\nu} \label{condec}
	\eeq
	where 
	\beq
	N_{\alpha\mu\nu}=\frac{1}{2}(Q_{\mu\nu\alpha}+Q_{\nu\alpha\mu}-Q_{\alpha\mu\nu}) -(S_{\alpha\mu\nu}+S_{\alpha\nu\mu}-S_{\mu\nu\alpha}) \label{l}
	\eeq
	is the so-called distortion tensor and $\widetilde{\Gamma}^{\lambda}_{\;\;\;\mu\nu}$ is the usual Levi-Civita connection given by
	\begin{equation}
	\widetilde{\Gamma}^{\lambda}_{\;\;\;\mu\nu}:=\frac{1}{2}g^{\alpha\lambda}(\partial_{\mu}g_{\nu\alpha}+\partial_{\nu}g_{\alpha\mu}-\partial_{\alpha}g_{\mu\nu})
	\end{equation}
From now on, all Riemannian quantities (i.e. evaluated with respect to the Levi-Civita connection) will be denoted by a tilde. We have now briefly developed the  geometric setup to be used for the rest of our analysis. For en extended exposure on the aspects of non-Riemannian geometry the reader is referred to \cite{eisenhart2012non}.
	\section{The sources of Metric-Affine Gravity (MAG)}
	\subsection{Canonical and Metrical Energy Momentum and Hypermomentum Tensors}
	As we have already pointed out in the MAG framework one starts  with the three independent fields $g_{ab}, \vtt^{b}$ and $\Gamma^{a}_{\;\; b}$. The field equations of a given MAG Theory are obtained by varying the total action independently with respect to those fields. Then, the variations of the matter part of the actions would be the sources of Gravity. Let $\mathcal{L}_{m}$ be the matter Lagrangian of the Theory. Then we have the following variations:
	\newline
		\underline{Canonical (Noether) Energy Momentum ($n-1$)-form}
	\beq
\Sigma_{a}:=	\frac{\delta \mathcal{L}_{m}}{\delta \vtt^{a}}
\eeq
	\underline{	Metrical (Hilbert) Energy Momentum $n$-form}
\beq
\sigma^{ab}:=2 \frac{\delta \mathcal{L}_{m}}{\delta g_{ab}}	
\eeq
	\underline{	Hypermomentum ($n-1$)-form}
	\beq
\Delta_{a}^{\;\; b}:=-2\frac{\delta \mathcal{L}_{m}}{\delta \Gamma^{a}_{\;\; b}}	
\eeq
Therefore, the three sources of MAG are the canonical, the metrical and the hypermomentum currents of matter \cite{hehl1976hypermomentum,hehl1978hypermomentum}. Note that the hypermomentum can be split into its three  irreducible pieces of spin, dilation and shear according to
\beq
\Delta_{ab}=\tau_{ab}+\frac{1}{n}\Delta g_{ab}+\hat{\Delta}_{ab}
\eeq
where $\tau_{ab}:=\Delta_{[ab]}$ is the spin part $\Delta:=\Delta_{cd}g^{cd}$ the dilation (trace) and $\hat{\Delta}_{ab}$ the shear (symmetric traceless part). Of course, the physical role of spin and dilation is well known. The most elusive so far has been the  role of shear. There have been some interesting early attempts to connect it to the hadronic properties of matter \cite{hehl1997ahadronic} but this connection is not totally clear. For a recent study on the role of shear hypermomentum in Cosmology, see \cite{iosifidis2020non}.
 In a holonomic frame  the above canonical, metrical energy momentum and hypermomentum tensors read 
 	\beq
 t^{\mu}_{\;\; c}=\frac{1}{\sqrt{-g}}\frac{\delta S_{M}}{\delta e_{\mu}^{\;\; c}}
 \eeq
\beq
T^{\alpha\beta}:=+\frac{2}{\sqrt{-g}}\frac{\delta(\sqrt{-g} \mathcal{L}_{M})}{\delta g_{\alpha\beta}}
\eeq
\beq
\Delta_{\lambda}^{\;\;\;\mu\nu}:= -\frac{2}{\sqrt{-g}}\frac{\delta ( \sqrt{-g} \mathcal{L}_{M})}{\delta \Gamma^{\lambda}_{\;\;\;\mu\nu}}
\eeq
where we have made the identifications $T_{ab}=-\star \sigma_{ab}$, $t_{ab}=e_{b}\rf \star\Sigma_{a}$ and $
\Delta_{a}^{\;\; b d}:= g^{cd} e_{c}\rf \star \Delta_{a}^{\;\; b}
$
in order to extract the tensor components. Next we discuss the conservation laws these sources must obey.
	
	\subsection{Conservation Laws}
As we have discussed in the previous section, the sources of MAG are the canonical  energy momentum tensor along with the hypermomentum current of matter. Of course there is also the metrical energy momentum tensor but this can be seen as a byproduct of the latter two \cite{hehl1995metric}. These sources are not quite independent and obey certain conservation laws as generalized versions of the energy momentum conservation in GR. Indeed, working in the exterior calculus language, the diffeomorphism invariance of the matter part of the action gives, on-shell \cite{hehl1995metric,gronwald1997metric}\footnote{Translated here to our conventions.}
	\beq
D \Sigma_{a}=\frac{1}{2}(e_{a}\rf R^{b}_{\;\; c})\w \Delta_{b}^{\;\; c}-\frac{1}{2}(e_{a}\rf \mathcal{Q}_{bc})\sigma^{bc}+(e_{a}\rf \mathcal{T}^{b})\w \Sigma_{b} \label{MCL1}
\eeq  
In addition, we now also have $GL(n,R)$ invariance which when applied to the matter part implies (still on-shell)
\beq
D \Delta_{a}^{\;\; b}=2(\vtt^{b}\w \Sigma_{a}-\sigma^{b}_{\;\; a}) \label{MCL2}
\eeq
The above two equations give the set of conservation laws which have to be obeyed by the matter sources of MAG. Switching to a holonomic frame, they read \cite{obukhov2013conservation,damianos2020cosmological}
		\beq
	t^{\mu}_{\;\;\lambda}
	= T^{\mu}_{\;\;\lambda}-\frac{1}{2 \sqrt{-g}}\hat{\nabla}_{\nu}(\sqrt{-g}\Delta_{\lambda}^{\;\;\mu\nu}) \label{cc1}
	\eeq
	\beq
	\frac{1}{\sqrt{-g}}\hat{\nabla}_{\mu}(\sqrt{-g}t^{\mu}_{\;\;\alpha})=-\frac{1}{2} \Delta^{\lambda\mu\nu}R_{\lambda\mu\nu\alpha}+\frac{1}{2}Q_{\alpha\mu\nu}T^{\mu\nu}+2 S_{\alpha\mu\nu}t^{\mu\nu} \label{cc2}
	\eeq
	where 
	\beq
	\hat{\nabla}_{\nu}:=2S_{\nu}-\nabla_{\nu}
	\eeq
The above are the MAG conservation laws in the holonomic description.
	For the most part we will be using this coordinate based formalism (holonomic), which makes things more transparent, but we will give the construction equations of our Perfect Hyperfluid Model in both languages. Again, the set of (\ref{cc1})-($\ref{cc2}$) or equivalently ($\ref{MCL1}$)-(\ref{MCL2}) has to be obeyed by all matter types of MAG and will be crucial in developing our Perfect Hyperfluid Model.
	
	\section{ Perfect Hyperfluid: Foundation}
	We shall now define the Perfect Hyperfluid as a direct generalization of the Perfect fluid of GR, where now the microstructure of matter (hypermomentum) is also taken into account. Our definition for Perfect Hyperfluid goes as follows.
	\begin{mydef}
	 Let $t_{\mu\nu}$, $T_{\mu\nu}$ and $\Delta_{\alpha\mu\nu}$ represent the canonical and metrical energy momentum tensors and $\Delta_{\alpha\mu\nu}$ the hypermomentum of the fluid. We define the Perfect Hyperfluid as exactly this fluid whose associated energy tensors ($t,T, \Delta$) respect spatial isotropy\footnote{In other words, following Weinberg's definition \cite{weinberg1972gravitation}, in our case ' A perfect hyperfluid is defined as this fluid where there exists a velocity $\vec{v}$ such that an observer moving with this velocity will see their surroundings as isotropic'. I am very grateful to Jose Beltran Jimenez for bringing this definition to my attention.  }. Mathematically,  we demand a vanishing Lie derivative along the spatial slices for each (see also \cite{tsamparlis1979cosmological}), i.e.
	\beq
\Lie_{\xi}t_{\mu\nu}=0\;\;, \;\; \Lie_{\xi}T_{\mu\nu}=0\;\;, \;\;   \Lie_{\xi}\Delta_{\alpha\mu\nu}=0
\eeq
	\end{mydef}
This implies that both the canonical and the metrical energy momentum tensors would have the perfect fluid form\footnote{Here, as usual, we consider a normalized velocity field $u_{\mu}u^{\mu}=-1$ and perform an $1+(n-1)$ spacetime split with the projection tensor $h_{\mu\nu}=g_{\mu\nu}+u_{\mu}u_{\nu}$.}
\beq
t_{\mu\nu}=\rho_{c}u_{\mu}u_{\nu}+p_{c} h_{\mu\nu} \label{canonical}
\eeq
\beq
T_{\mu\nu}=\rho u_{\mu}u_{\nu}+p h_{\mu\nu} \label{metrical}
\eeq
where $\rho_{c},p_{c}$ are the density and pressure associated with the canonical part and $\rho,p$ the usual ones associated to $T_{\mu\nu}$. In addition, demanding only spatial isotropy  the hypermomentum takes the covariant form \cite{damianos2020cosmological}	
	\beq
	\Delta_{\alpha\mu\nu}^{(n)}=\phi h_{\mu\alpha}u_{\nu}+\chi h_{\nu\alpha}u_{\mu}+\psi u_{\alpha}h_{\mu\nu}+\omega u_{\alpha}u_{\mu} u_{\nu}+\delta_{n,4}\epsilon_{\alpha\mu\nu\kappa}u^{\kappa}\zeta \label{Dform}
	\eeq
	this is the most general covariant form of a type ($0,3$) tensor respecting isotropy. In the above $\delta_{n,4}$ is the Kronecker's delta.  Note that since we have not imposed homogeneity here, in contrast to \cite{damianos2020cosmological}, all the functions of the set $V=\{\rho_{c},p_{c},...\omega, \zeta \}$ will be generic spacetime functions i.e. $\psi=\psi(x^{\alpha})=\psi(t,\vec{x})$ etc. In a covariant fashion, these read
	\beq
\omega=-u^{\alpha}u^{\mu}u^{\nu}\Delta_{\alpha\mu\nu}
\eeq
\beq
\phi=-\frac{1}{(n-1)}h^{\alpha\mu}u^{\nu}\Delta_{\alpha\mu\nu}
\eeq
\beq
\chi=-\frac{1}{(n-1)}h^{\alpha\nu}u^{\mu}\Delta_{\alpha\mu\nu}
\eeq
\beq
\psi=-\frac{1}{(n-1)}h^{\mu\nu}u^{\alpha}\Delta_{\alpha\mu\nu}
\eeq
\beq
\zeta=+\frac{1}{6}\epsilon^{\alpha\mu\nu\lambda}\Delta_{\alpha\mu\nu}u_{\lambda}\delta_{n,4}
\eeq
These are the $5$ material variables describing the hypermomentum part of the fluid. These five fields are then rearranged in a certain way and provide the spin, dilation and shear parts according to
	\beq
	\Delta_{[\alpha\mu]\nu}=(\psi-\chi)u_{[\alpha}h_{\mu]\nu}+\delta_{n,4}\epsilon_{\alpha\mu\nu\kappa}u^{\kappa}\zeta
	\eeq
	\beq
	\Delta_{\nu}:=\Delta_{\alpha\mu\nu}g^{\alpha\mu}=\Big[ (n-1) \phi -\omega\Big] u_{\nu} \label{dil}
	\eeq
	\beq
	\breve{\Delta}_{\alpha\mu\nu}=\Delta_{(\alpha\mu)\nu}-\frac{1}{n}g_{\alpha\mu}\Delta_{\nu} =\frac{(\phi+\omega)}{n}\Big[ h_{\alpha\mu}+(n-1)u_{\alpha}u_{\mu} \Big] u_{\nu} +(\psi +\chi)u_{(\mu}h_{\alpha)\nu}
	\eeq
	Let us stress again that all the fields appearing above are generic spacetime functions.

	\subsection{Conservation Laws of the Perfect Hyperfluid}
	We are now in a position to derive the conservation laws that hold true for our Perfect Hyperfluid Model. As we discussed earlier, in this case the canonical energy momentum tensor has the perfect fluid form and it is therefore symmetric as seen from ($\ref{canonical}$). Then, for any symmetric rank-$2$ tensor $C_{\mu\nu}$ we have the identity
	\beq
	-\frac{1}{\sqrt{-g}}\hat{\nabla}_{\mu}(\sqrt{-g}C^{\mu}_{\;\;\nu})=\tilde{\nabla}_{\mu}C^{\mu}_{\;\; \nu}-\frac{1}{2}Q_{\nu\alpha\beta}C^{\alpha\beta}-2 S_{\nu\alpha\beta}C^{\alpha\beta}
	\eeq
	which can be proved trivially by expanding the left-hand side. Applying this to the canonical energy momentum tensor, the conservation law $(\ref{cc2})$ reads
		\beq
	\tilde{\nabla}_{\mu}t^{\mu}_{\;\; \alpha}=\frac{1}{2}\Delta^{\lambda\mu\nu}R_{\lambda\mu\nu\alpha}+\frac{1}{2}Q_{\alpha\mu\nu}(t^{\mu\nu}-T^{\mu\nu}) \label{CL1}
	\eeq
	which is also supplemented by 
		\beq
	t^{\mu}_{\;\;\lambda}
	= T^{\mu}_{\;\;\lambda}-\frac{1}{2 \sqrt{-g}}\hat{\nabla}_{\nu}(\sqrt{-g}\Delta_{\lambda}^{\;\;\mu\nu}) \label{CL2}
	\eeq
	From the first one above we see that the difference of the canonical and the metrical tensors couples directly to spacetime non-metricity. Of course in the case of the Perfect Cosmological Hyperfluid of \cite{damianos2020cosmological} this term drops out since in that case the latter two tensors coincide. Note also that the fact that in general the metrical is not the same with the canonical implies that the hypermomentum is not conserved, as seen from ($\ref{CL2}$). In addition, as it is obvious from the above relations one may eliminate the difference of the two energy tensors from ($\ref{CL1}$) by using $(\ref{CL2})$. Then we get the alternative expression
		\beq
	\tilde{\nabla}_{\mu}t^{\mu}_{\;\; \alpha}=\frac{1}{2}\Delta^{\lambda\mu\nu}R_{\lambda\mu\nu\alpha}-\frac{1}{4 \sqrt{-g}}Q_{\alpha\mu}^{\;\;\;\;\lambda}\hat{\nabla}_{\nu}(\sqrt{-g}\Delta_{\lambda}^{\;\;\;\mu\nu})
	\eeq
	The advantage of this last expression is that it involves only the canonical and the hypermomentum tensors and not the metrical which is a byproduct of the latter two, as seen from ($\ref{CL2}$). Let us collect all the above results and postulate the Perfect Hyperfluid concept along with its complete mathematical description.
	\begin{mydef}
The Perfect Hyperfluid: There exists a Perfect (ideal) Hyperfluid structure, carrying intrinsic hypermomentum, that generalizes the Perfect Fluid concept. The description of the Perfect Hyperfluid is given by the energy tensors
	\beq
	t_{\mu\nu}=\rho_{c}u_{\mu}u_{\nu}+p_{c} h_{\mu\nu} \label{canonical}
	\eeq
	\beq
	T_{\mu\nu}=\rho u_{\mu}u_{\nu}+p h_{\mu\nu} \label{metrical}
	\eeq
		\beq
	\Delta_{\alpha\mu\nu}^{(n)}=\phi h_{\mu\alpha}u_{\nu}+\chi h_{\nu\alpha}u_{\mu}+\psi u_{\alpha}h_{\mu\nu}+\omega u_{\alpha}u_{\mu} u_{\nu}+\delta_{n,4}\epsilon_{\alpha\mu\nu\kappa}u^{\kappa}\zeta \label{Dform}
	\eeq
subject to the conservation laws
			\beq
	\tilde{\nabla}_{\mu}t^{\mu}_{\;\; \alpha}=\frac{1}{2}\Delta^{\lambda\mu\nu}R_{\lambda\mu\nu\alpha}+\frac{1}{2}Q_{\alpha\mu\nu}(t^{\mu\nu}-T^{\mu\nu}) \label{CL1}
	\eeq
	\beq 
	t^{\mu}_{\;\;\lambda}
	= T^{\mu}_{\;\;\lambda}-\frac{1}{2 \sqrt{-g}}\hat{\nabla}_{\nu}(\sqrt{-g}\Delta_{\lambda}^{\;\;\mu\nu}) \label{CL2}
	\eeq
	providing a direct generalization of the Perfect Fluid continuum when the intrinsic characteristics of matter (i.e. $\Delta_{\alpha\mu\nu}$) are also taken into account. 
		\end{mydef}
	\textbf{Comment:} As seen from the above, the complete description of the Perfect Hyperfluid is given by  the set of $9$ spacetime functions $\{\rho,\rho_{c},p,p_{c},\phi,\chi,\psi,\omega,\zeta\}$ along with its associated velocity field $u$.
	\subsection{Exterior Calculus representation of Perfect Hyperfluid}
	
	For completeness we will give the forms of energy tensors of the Perfect Hyperfluid (MAG sources) in the language of exterior differential forms, which is of great use in MAG. It can be easily seen that the hypermomentum expression ($\ref{Dform}$), in the language of differential forms, translates to the hypermomentum ($n-1$) form 
	\beq
	\Delta_{a}^{\;\; b}=\delta_{a}^{b}\phi u+\chi (e^{b}\rf \star u)\star \vtt_{a}+\psi (e_{a}\rf \star u)\star \vtt^{b}+(e_{a}\rf \star u) (e^{b}\rf \star u) \bar{\omega} u-3!\vtt_{a}\w \vtt^{b}\w Z \delta_{n,4} \label{Dform2}
	\eeq
	where $\phi,\chi,\psi$ and $\bar{\omega}=\omega+\phi+\chi+\psi$ are $0$-forms and $Z=\zeta_{a}\vtt^{a}=\zeta u_{a}\vtt^{a}=\zeta \star u$ is an $1$-form. The associated canonical ($n-1$)-from and metrical $n$-form currents are also extracted rather trivially and read
	\beq
	\Sigma_{a}=(\bar{\rho}+\bar{p})(e_{a}\rf\star u)u+\bar{p} \star \vtt_{a} \label{tform}
	\eeq
		\beq
	\sigma_{ab}=\mu \Big( (\rho+p)u_{a}u_{b}+pg_{ab}\Big) \label{Tform}
	\eeq
	 respectively. In the above with have denoted $\rho_{c}=\bar{p}$ and $p_{c}=\bar{p}$ in order to avoid confusion with the anholonomic index $c$. In addition, we have considered the flow ($n-1$)-form (see for instance \cite{obukhov1996model}) $u$ with its dual giving us the velocity field
	 \beq
	 \star u:=u_{a} \vtt^{a}
	 \eeq
	and the normalization
	\beq
u \w \star u=\mu	
	\eeq
	With this we may re-express ($\ref{Dform2}$) in the more transparent form
		\beq
	\Delta_{a}^{\;\; b}=\delta_{a}^{b}\phi u+\chi u^{b}(\star \vtt_{a})+\psi u_{a}(\star \vtt^{b})+u_{a}u^{b} \tilde{\omega} u-3! \zeta\vtt_{a}\w \vtt^{b}\w (\star u) \delta_{n,4}
	\eeq
	The latter expression along with ($\ref{tform}$) and ($\ref{Tform}$) represent the material sources of the Perfect (Ideal) Hyperfluid in the language of differential forms and are subject to the conservation laws ($\ref{MCL1}$) and $(\ref{MCL2})$.

	\section{Theories with $\mathcal{L}_{G}=\mathcal{L}_{G}(R^{\lambda}_{\;\;\;\alpha\beta\gamma})$}
	In the special case of Theories whose Gravitational part is constructed only by the Riemann tensor and its contractions we have a very important restriction on the Hyperfluid sector. Indeed, as can be trivially checked in this case, the Riemann tensor is invariant under special projective transformations of the form\footnote{Also known as $\lambda$-transformations.}
	\beq
	\Gamma^{\lambda}_{\;\;\;\mu\nu} \longrightarrow \bar{\Gamma}^{\lambda}_{\;\;\;\mu\nu}=\Gamma^{\lambda}_{\;\;\;\mu\nu} +\delta^{\lambda}_{\mu}\partial_{\nu}\lambda
	\eeq 
	\beq
	R^{\lambda}_{\;\;\mu\nu\alpha} \longrightarrow 	\bar{R}^{\lambda}_{\;\;\mu\nu\alpha}=R^{\lambda}_{\;\;\mu\nu\alpha}
	\eeq
	Of course, all  contractions of the Riemann tensor will also respect this symmetry. The above fact has a great impact on the hypermomentum sources. Indeed, since this symmetry have to be respected from the matter part as well, we get the constraint (see for instance \cite{iosifidis2019scale})
	\beq
	\partial_{\nu}(\sqrt{-g}\Delta^{\nu})=0
	\eeq
	that is, the dilation current $\Delta^{\mu}:=\Delta_{\lambda}^{\;\;\lambda\nu}$ must be conserved. Equivalently, taking the contraction of $(\ref{CL2})$ the above constraint translates to
	\beq
	t=T
	\eeq
	namely the traces of the canonical and the metrical tensors must coincide. Note that up to now we have made no assumption on the matter content of the Theory. If we apply the above result to our Perfect Hyperfluid model, given the forms ($\ref{canonical}$) and ($\ref{metrical}$) it follows that
	\beq
-\rho_{c}+(n-1)p_{c}=-\rho+(n-1) p	
	\eeq 
	In addition, assuming that both perfect fluid components (metrical and canonical) are barotropic, that is $p_{c}=w_{c}\rho_{c}$ and $p=w\rho$, we have
	\beq
\rho_{c}=\left(\frac{1-(n-1)w}{1-(n-1)w_{c}}\right)\rho	 \label{rr}
	\eeq
	From the above discussion it is now clear that if the barotropic components of the canonical and the metrical are of the same kind, in the sense that $w=w_{c}$ we immediately have that $p=p_{c}$ and from ($\ref{rr}$) it follows that $\rho=\rho_{c}$ as well. Conversely, on the (logical!) assumption that the  associated densities are equal (i.e. $\rho=\rho_{c}$) we also have that $p=p_{c}$ from the above equation. Either way the end result is that both the pressures and the densities would be identical and consequently $t_{\mu\nu}\equiv T_{\mu\nu}$. We therefore see how the Perfect Cosmological Hyperfluid Model of \cite{damianos2020cosmological} is embedded in our general construction here. Given that the Gravitation part is built only from the Riemann tensor and its contractions, the Perfect Hyperfluid of \cite{damianos2020cosmological} is the one for which the barotropic fluid components are of the same kind $w=w_{c}$ or alternatively the canonical and metrical densities coincide. Recall that in this case the conservation laws take the form
	\beq
	\widetilde{\nabla}_{\mu}T^{\mu}_{\;\;\nu}=\frac{1}{2} \Delta^{\alpha\beta\gamma}R_{\alpha\beta\gamma\nu} 
	\eeq
	\beq
	\hat{\nabla}_{\nu}\Big( \sqrt{-g}\Delta_{\lambda}^{\;\;\;\mu\nu}\Big)=0 \;\;, \;\;\; t_{\mu\nu}=T_{\mu\nu}
	\eeq	
	For this reason we shall call this Model the \underline{Hypermomentum Conserving Perfect  Hyperfluid}. The above considerations were rather general with no assumption about the underlying spacetime. If we consider a Cosmological setting the fluid obeying the above two conservation laws (that is, the fluid in \cite{damianos2020cosmological})  	 will be called the \underline{Hypermomentum Preserving Perfect Cosmological Hyperfluid}.

	\section{Cosmological Application of the Perfect Hyperfluid}
	Let uf now see an immediate application of our Fluid model. If we now impose also homogeneity and consider and FLRW Universe all the variables, of the hyperfluid, would be time dependent only and in this case our conservation laws ($\ref{CL1}$) and ($\ref{CL2}$) boil down to\footnote{As usual, the dot denotes time derivative.}
\beq
\Big[ \dot{\rho}+(n-1)H(\rho+p) \Big] u_{\nu}+(\rho +p)u^{\mu}\widetilde{\nabla}_{\mu}u_{\nu}=\frac{1}{2}u^{\mu}(\phi \widehat{R}_{\mu\nu}+\chi R_{\mu\nu}+\psi \breve{R}_{\mu\nu})+\frac{1}{2}u_{\nu}\Big[ (\rho_{c}-\rho)C+(p_{c}-p)A \Big] \label{cl1}
\eeq
\begin{gather}
-\delta^{\mu}_{\lambda}    \frac{\partial_{\nu}(\sqrt{-g}\phi u^{\nu})}{\sqrt{-g}}-u^{\mu}u_{\lambda}      \frac{\partial_{\nu}\Big(\sqrt{-g}(\phi+\chi +\psi +\omega) u^{\nu}\Big)}{\sqrt{-g}}
\nonumber \\
+\left[ \Big(2 S_{\lambda}+\frac{Q_{\lambda}}{2}\Big)u^{\mu}-\nabla_{\lambda}u^{\mu} \right]\chi +\left[ \Big(2 S^{\mu}+\frac{Q^{\mu}}{2}-\tilde{Q}^{\mu}\Big)u_{\lambda}-g^{\mu\nu}\nabla_{\nu}u_{\lambda}\right]\psi
\nonumber \\
+ u^{\mu}u_{\lambda}(\dot{\chi}+\dot{\psi}) -(\phi+\chi+\psi+\omega)(\dot{u}^{\mu}u_{\lambda}+u^{\mu}\dot{u}_{\lambda}) 
=2(\rho -\rho_{c})u_{\lambda}u^{\mu}+2 (p-p_{c})h_{\lambda}^{\;\;\mu}   \label{conl2}
\end{gather}
	The above equations contain the full dynamics of the generalized Perfect Cosmological Hyperfluid. Let us highlight again that the degrees of freedom in this case are\footnote{Obviously two degrees of freedom come from $\rho$, $p$ another two from $\rho_{c}$, $p_{c}$ and the rest  are the five hypermomentum variables.} $2+2+5=9$. However, due to the high symmetry of FLRW spacetime, the above conservation laws only provide $1+2$ evolution equations. This is no surprise since a similar situation appears in the case of the Perfect Fluid of GR where the continuity equation only gives the evolution for $\rho$ and p is usually assumed to satisfy an equation of state $p=w\rho$ such that the system becomes complete. The same behaviour we also expect here, so we will need to provide two barotropic indices for the metrical and canonical parts according to
	\beq
	p=w \rho \;\;, \;\;\; p_{c}=w_{c} \rho_{c} \label{pr}
	\eeq	
	In addition, there should be one equation relating the above $4$ functions (similar to ($\ref{rr}$)) and we should also have  three equations of state among the hypermomentum variables, in order to have a completely determined dynamics.  In general, the three equations of state among the hypermomentum variables would have the form
	\beq
	F_{A}(\phi,\chi,\psi,\omega,\zeta)=0\;, \;\;\; A=1,2,3
	\eeq
	To further restrict the above possibility it would be most natural to associate equations of state among the different parts of hypermomentum (spin, dilation and shear) as was implemented in \cite{iosifidis2020non}. The exact values of these equations of state would characterize the nature of the hyperfluid (see also \cite{damianos2020cosmological} where such equations of state are derived). We should also note that it would be possible to have equations of state that mix up the perfect fluid with the hypermomentum parts. Then, one needs $6$ independent equations of state, whose generic form would read
	\beq
F_{I}(\rho, p, \rho_{c}, p_{c},\phi,\chi,\psi,\omega,\zeta)=0 \;\; , \;\;\; I=1,2,...,6
	\eeq   
	As a result, the Perfect Cosmological Hyperfluid lies in the intersection of the aforementioned $6$ hypersurfaces. However, we find the latter mixing possibility very unlikely, though it may be true for some types of generalized hyperfluids. In any case, the three conservation laws for any type of Cosmological  Hyperfluids, can be extracted from $(\ref{cl1})$ and $(\ref{conl2})$ by first contracting the former with $u^{\mu}$ and take from the latter one time the $ij$-components and another the $00$ to arrive at
	\beq
\dot{\rho}_{c}+(n-1)H(\rho_{c}+p_{c})	=-\frac{1}{2}u^{\mu}u^{\nu}(\chi R_{\mu\nu}+\psi \breve{R}_{\mu\nu})+\frac{1}{2}(\rho_{c}-\rho)C+\frac{1}{2}(p_{c}-p)A \label{cont}
	\eeq	
	\beq
\dot{\phi}+(n-1)H \phi +H(\chi +\psi) +\psi X- \chi Y=2 (p_{c}-p) 	 \label{hyper1}
	\eeq
	\beq
	\dot{\omega}+(n-1)H(\chi+\psi+\omega)+(n-1)( \psi X-\chi Y)=2(\rho_{c}-\rho) \label{hyper2}
	\eeq
Additionally, one could take the trace of ($\ref{conl2}$) to arrive at
	\beq
	(n-1)\dot{\phi}-\dot{\omega}+(n-1)H \Big[ (n-1)\phi-\omega \Big]=2\Big[ (\rho-\rho_{c})-(n-1)(p-p_{c})  \Big] \label{dil}
	\eeq
	However, as expected, this gives no further information since it is easily seen that the latter is equal to\footnote{As is also apparent from $(\ref{dil})$ the dilation current is conserved only when $(\rho-\rho_{c})=(n-1)(p-p_{c})$. This condition is always true for frame rescaling invariant Theories \cite{iosifidis2019scale}. } $(n-1)(\ref{hyper1})-(\ref{hyper2})$. Therefore, as we have already mentioned, the full dynamics of the Perfect Hyperfluid is contained in the three equations $(\ref{cont})$, $(\ref{hyper1})$ and $(\ref{hyper2})$. We should emphasize again that the latter equations are fairly general and hold true regardless of the equations of state to be imposed on the hyperfluid variables. For any Metric-Affine Cosmology, the evolution equations for the sources are the aforementioned three.

	Let us now organize the above ideas according to the following classifications. We start with the most general case and subsequently specialize.

	\subsection{Cosmological Hyperfluids Classification}
	
Below we classify some characteristic cases of the Perfect Cosmological Hyperfluids.
\begin{mydef} A \underline{generalized Perfect Cosmological Hyperfluid} consists of a set of energy tensors given by the expressions ($\ref{canonical}$), ($\ref{metrical}$) and $(\ref{Dform})$  subject to the conservation laws 
	\beq
	\tilde{\nabla}_{\mu}t^{\mu}_{\;\; \alpha}=\frac{1}{2}\Delta^{\lambda\mu\nu}R_{\lambda\mu\nu\alpha}+\frac{1}{2}Q_{\alpha\mu\nu}(t^{\mu\nu}-T^{\mu\nu}) \label{CLN1}
	\eeq
	\beq 
	t^{\mu}_{\;\;\lambda}
	= T^{\mu}_{\;\;\lambda}-\frac{1}{2 \sqrt{-g}}\hat{\nabla}_{\nu}(\sqrt{-g}\Delta_{\lambda}^{\;\;\mu\nu}) \label{CLN2}
	\eeq
	In addition, its $9$ material sources are related to one another by the $6$ generalized equations of state
	\beq
	F_{I}(\phi,\chi,\psi,\omega,\zeta,\rho,p,\rho_{c},p_{c})=0 \;, \;\;\; 	 I=1,2,...,6 \label{F}
	\eeq
	\end{mydef}

\begin{mydef}A \underline{Barotropic Perfect Cosmological Hyperfluid} represents a special case of the above for which the equations of state ($\ref{F}$) have the barotropic form
	\beq
	\sum_{i=1}^{9}	a_{i}^{I}X_{i}=0\;, I=1,2,...,6 \;,\;\; i=1,2,...,9
	\eeq
	where not all  $a_{i}^{I}$'s  are zero and we have also collectively denoted $X_{i}=\{ \rho_{c},p_{c},\rho,p,\phi,\chi,\psi,\omega,\zeta\}	$. 
	\end{mydef}
\begin{mydef} A \underline{Decoupled Barotropic Perfect Cosmological Hyperfluid} is the one for which the barotropic equations among its constituents have the following  forms
\beq
p=w \rho, \;\; p_{c}=w_{c}p_{c}\;, \;\; , \rho_{c}=\bar{w}\rho
\eeq
	\beq
\sum_{i=1}^{6}	a_{i}^{I}Y_{i}=0\;, I=1,2,...,6 \;,\;\; i=1,2,...,5
\eeq
where not all  $a_{i}^{I}$'s  are zero and we have collectively denoted $Y_{i}=\{\phi,\chi,\psi,\omega,\zeta\}	$. 

\textbf{Comment:} The Perfect Hypermomentum Preserving Cosmological Hyperfluid (see previous section) represents a special case of the latter with $\rho_{c}=\rho$ and $p_{c}=p$.
\end{mydef}

	\section{Discussion}
	We have constructed a straight generalization of the familiar Perfect Fluid notion, encompassing now the microscopic characteristics of matter as well. We call this extended fluid notion a Perfect (Ideal) Hyperfluid. The latter is described by the three tensors of the canonical, metrical and hypermomentum currents, all of which have an isotropic form (see eqn's  $(\ref{canonical})$-$(\ref{Dform})$) and are subject to the MAG conservation laws.
As soon as the intrinsic properties of the fluid are neglected ($\Delta_{\alpha\mu\nu}=0$) we arrive at the classical Perfect Fluid model as a limiting case, as expected. The description of the generalized Perfect Hyperfluid is given by $9$ spacetime functions along with the fluids' velocity flow $u$. The specific form of these $9$ functions will characterize the fluid under consideration.

It should be noted that even though our generalized Perfect Hyperfluid construction fits most naturally in a Metric-Affine Gravity approach, this is by no means the only place it can find applications. Indeed, our general construction here can just as well be applied to all Theories that represent special cases of MAG. For instance, Einstein-Cartan \cite{trautman2006einstein} or more generalized torsionful Theories (like Poincare Gravity \cite{hayashi1980gravity}), non-metric torsionless Theories, and also to all teleparallel Theories such as metric \cite{aldrovandi2012teleparallel}, symmetric \cite{nester1998symmetric,jimenez2018teleparallel} and generalized teleparallelism \cite{jimenez2020general}. Of course  the list could go on and on.  In general we expect the Perfect Hyperfluid to describe matter configurations to all Gravity Theories exhibiting a non-trivial connection.  

Furthermore, the Perfect Hyperfluid could find interesting applications in the Theory of materials with microstructure  and  elasticity \cite{mindlin1963microstructure}. This is plausible since the very concept of hypermomentum has a close analogy with a notion appearing in materials with microstructure, which is known as hyper-stress \cite{gronwald1997stress}.  Lastly, probably the most obvious application of our hyperfluid structure would be to use it in order to study the full quadratic MAG Theory in a Cosmological setting and in the presence of the latter fluid configuration. This is currently under investigation.

\section{Acknowledgments} I would like to thank Jose Beltran Jimenez and Tomi Sebastian Koivisto for some useful discussions. 	This research is co-financed by Greece and the European Union (European Social Fund- ESF) through the
Operational Programme 'Human Resources Development, Education and Lifelong Learning' in the context
of the project “Reinforcement of Postdoctoral Researchers - 2
nd Cycle” (MIS-5033021), implemented by the
State Scholarships Foundation (IKY).

\bibliographystyle{unsrt}
\bibliography{ref}

\end{document}